# *OntoCat:* Automatically categorizing knowledge in API Documentation

Niraj Kumar
University of California, Davis
nuckumar@ucdavis.edu

Premkumar Devanbu
University of California, Davis
ptdevanbu@ucdavis.edu

## ABSTRACT

Most application development happens in the context of complex APIs; reference documentation for APIs has grown tremendously in variety, complexity, and volume, and can be difficult to navigate. There is a growing need to develop well-organized ways to access the knowledge latent in the documentation; several research efforts deal with the organization (ontology) of API-related knowledge. Extensive knowledge-engineering work, supported by a rigorous qualitative analysis, by Maalej & Robillard [3] has identified a useful taxonomy of API knowledge. Based on this taxonomy, we introduce a domain independent technique to extract the knowledge types from the given API reference documentation. Our system, *OntoCat,* introduces total nine different features and their semantic and statistical combinations to classify the different knowledge types. We tested *OntoCat* on python API reference documentation. Our experimental results show the effectiveness of the system and opens the scope of probably related research areas (i.e., user behavior, documentation quality, etc.).

## 1. INTRODUCTION

Modern applications are typically based on application platforms such as Android, .NET and Django. These platforms provide a rich, complex, variety of services, accessed via application programmer interfaces (APIs). Ideally, an API represents a clear, unambiguous way for an application developer to use services provided by the software platform to provide features customers desire. In practice, developers require a great deal of different types of knowledge in order use the APIs correctly. Thus a method like `writeToSocket()` might appear straightforward, but in fact might involve subtleties: *when can it be called? What can kind of socket can be written to? What if the network is unavailable?* and so on. To gain this type of needed knowledge, developers have to turn to API documentation.

API documentation can run into thousands of pages, describing the classes and methods that developers use to access these services. Developers using these APIs face the challenge of finding needed information in this trove. As noted by [3], the volume can be quite imposing: the documentation of JDK 6, taken together, is about 6 times the length of Tolstoy's "War and Peace".

These documents are usually indexed by API element name, where each document specifically provides information about an element (class, method, etc.). For example, in the case of Python, there is a set of pages, where each of them covers one API out of many, and each is broken up by class and method. However, this kind of indexing doesn't necessarily help developers find all the types of knowledge they might need; this leads to a lot of wasted time and effort. Thus, despite extensive (several thousand pages of documentation) associated with the Android API, there are over 800,000 questions in StackOverflow, tagged with "Android"[1]. This is not a problem peculiar to Android; there are over 425,000 questions tagged "iOS" and over 100,000 tagged "Django". Clearly, there is room for improvement in API documentation.

There has been considerable research into the kinds of knowledge that developers seek during development tasks [14,15,16]. Of particular relevance to us are the types of knowledge sought during API based development. There has been considerable interest in this topic, as well [3,12,13,17]. Most relevant to our concern here is the work of Maalej & Robillard [3]. In this work, the authors conduct an extensive content analysis [18] of both .NET and JDK documentation. The goal of their analysis was to develop a *carefully validated* taxonomy of the patterns of knowledge that obtain within API documentation. They argue that without such a taxonomy of the knowledge patterns, it would be difficult to understand what works well, what doesn't, and how one might improve API documentation.

They used rigorous qualitative methodology, based on grounded theory, to develop their knowledge classification. They found 12 categories of knowledge in API documentation (Section 1.2). Their goal, as described in the paper was to produce a classification scheme that was *"1) Reliable, in that different people consistently come to the same conclusion about the knowledge types contained in documentation unit. 2) Meaningful, listing knowledge types relevant to the practice of software development. 3) Fined-grained, providing more than just a few high-level categories"* [3].

We describe their taxonomy in more detail below, but first we present our contributions.

1. We developed a set of textual features, whose appearance in

---

[1] See http://stackoverflow.com/questions/tagged/android

2. the content is statistically highly indicative of the knowledge class of the context in which they appear.

3. Using these features, and a training set of several hundred pre-classified API document content units, we have trained an ensemble of classifiers, which we call *OntoCat*. To our knowledge *OntoCat* is the most comprehensive automated classifier of software document content, which appears to perform reasonably accurately, and is statistically far better than random.

The OntoCat knowledge classifier provides a scalable, and automated approach. Our work paves the way towards several important future goals:

1. Systematically analyzing API documents: by classifying the breadth and intensity of various types of knowledge content in documentation, one could try to identify which types are lacking and most needed, and try to improve documents.

2. Content extraction: using advanced statistical NLP methods [19], one could try to automatically extract and formally represent (for query answering and information retrieval) the knowledge within API documentation. Knowledge of different types could also be indexed and made available for ready use.

3. Studying & Using crowd-sourced documentation: The very large volumes of crowd-sourced documentation in StackOverflow etc., could be studied to understand the association of knowledge types with ratings, as well as a source for knowledge-types that may be lacking or scarce in formally created sources (e.g., manuals).

In next section, we present some details on the Taxonomy developed in [3].

## 2. Background & Motivation
Our work is motivated by the work of Maalej & Robillard [3] and we begin with this work.

### 2.1 Maalej-Robillard Classification
Maalej & Robillard [3] describes to our knowledge the most comprehensive, rigorous effort to identify the types of knowledge in API documentation. Their taxonomy classifies knowledge into 12 types, shown in Table 1. They used a rigorous, 4-phase process to identify these knowledge classes.

The process began with a manual content analysis of document elements by the two authors, using a grounded-theory and analytical approaches to identify knowledge types based on a carefully chosen (theory-driven) sampling of API document content. This phase produced a taxonomy. In the second step, Maalej & Robillard used the taxonomy to develop a coding guide [23] (to code document elements into knowledge classes) and trained 17 coders to manually process a random sample of document units "to assess whether each unit contained knowledge of the different types in (the) taxonomy" [3]. Each unit was independently rated by two human coders. Agreements and disagreements were carefully analyzed and dealt with; they report significant agreement between human coders. In Phase 3, they carefully resolved conflicted ratings, using a rigorous procedure design to maximize the chances of choosing the most likely rating. In all, knowledge classes of 5,574 individual units were manually assigned, resulting in total of around 11,000 ratings (average of 2 ratings per unit). The entire process was conducted with a great degree of openness & reflection, resulting in the knowledge classes shown in Table 1. There are 12 classes listed. All are fully described in their paper [3], and in fuller detail in the coding guide [23]. The table presents a brief description of each knowledge class, together with an illustrative document unit, drawn from Python documentation.

### 2.2 Why Classify Automatically?
In [3] the authors use this knowledge classification scheme to manually analyze a large sample of document elements from .NET and JDK. This manually annotated set was used to compare .NET and JDK document quality. The authors argue that this type of content-based analysis of API documentation (Section 7.2, [3]) can help a) *analyze and improve* the content API documentation b) *provide an organization or indexing scheme*; c) *provide a vocabulary* for discussions about the document and d) *help structure methods to evaluate the quality of documentation.*

**Table-1: Taxonomy of Knowledge Types**

| Knowledge Type | Description & Example |
|---|---|
| Functionality and Behavior | Describes the function of some aspect of the API. <br> *E.g., Replace special characters in string using the %xx escape. Letters, digits, and the characters '_.-' are never quoted.* |
| Concepts | Describes terms or concepts associated with elements of the API. <br> *E.g., Dbm objects behave like mappings (dictionaries), except that keys and values are always stored as bytes* |
| Directives | Describes prescribed/disallowed operations on API elements. <br> *E.g., A returned value should be interpreted as the name of a global variable. It should be the object's local name relative to its module;* |
| Purpose and Rationale | Explains the purpose, or design rationale, of an API element. <br> *E.g., This is primarily used for list subclasses, but may be used by other classes as long as they have ... (etc)* |
| Quality Attributes and Internal Aspects | Describes quality and/or non-functional aspects of an API's external behavior, or internal implementation. <br> *E.g., The dbm.ndbm module provides an interface to the Unix "(n)dbm" library.* |
| Control-Flow | Describes API event execution order, or causality relationships. <br> *E.g., eHeader's __new__ then creates the header instance, and calls its init method.* |
| Structure | Describes internal organization of compound API elements. <br> *E.g., BaseHeader also provides the following method, which is called by the email library code ....* |
| Patterns | Describes how to achieve specific goals using the API, or how to customize the behavior of an API element. <br> *E.g., to write bytes to stdout, use sys.stdout.buffer.write(b'abc').* |
| *Code Examples* | Examples of code that implement specific functionality, or designs. <br> *E.g., make_streams_binary(): sys.stdin = sys.stdin.detach() sys.stdout = sys.stdout.detach() Not* |
| Environment | Describes aspects surrounding API use, such as compatible versions, platforms, licensing issues, etc. <br> *E.g., This is not the version of the SQLite library.* |
| References | A pointer to external documentation, such as a hyperlink, or a reference, or a citation of some sort. <br> *E.g., See http://www.wsgi.org for more information about WSGI, and links to tutorials and other resources.* |
| Non-information | A document fragment that contains uninformative, or boiler-plate text. <br> *E.g., doctest prints a de.* |

However, thus far, there has no way to automatically classify API document content into these knowledge types; classification must be manually executed, and thus is not scalable. The above potential applications of document content classification cannot, therefore, be applied broadly. This motivates the need for automatic classification of the knowledge types associated with knowledge elements.

In addition to the points noted in [3], we were inspired by the enormous body of work in Automated Content Extraction [19], which have enabled modern question-answering systems [24] that have achieved several spectacular successes. Over the past several decades a range of different statistical & machine learning methods have been developed to analyze textual content, from topic analysis models [25] to distant supervision [26]. The goals range from identifying broad topics in documents, to extracting specific facts (such sports team statistics) from free from natural language text. These methods have grown highly sophisticated, and capable of identifying not just facts, but also sentiment, rhetorical stance, and social status. While topic analysis has found wide and sophisticated application in software engineering [27,28]; however, more modern content extraction methods, to our knowledge, have not found application over software documentation. We hypothesize that one reason for this might be the existence of a wide range of knowledge classes in software manuals and documents; we believe that automatically labeled knowledge units could then subsequently automatically processed to extract formal representations of these knowledge classes, which could then used to create a sort of "Watson for Code".

In the next section, we present the details of our approach.

## 3. Technical Approach

Automatic content extraction and classification methods are based broadly on supervised machine learning approaches: one uses a set of features, a training set, and a machine learning algorithm to learn the association between the features and the labels in the training set.

For our training set, we used a dataset constructed by annotating Python API documents. This manually annotated dataset contains a total of 5,737 elements. At the time of research, we had access to a total of 1,864 annotated documents that we used for training, and testing an automatic classifier.

In the following, we describe the technical details of OntoCat's automatic classification approach, beginning with our choice of features.

### 3.1 Summary
Given a document fragment to be classified, we use 5 broad categories of features. These features are all well-founded in current literature in information retrieval and automated content extraction [30,31,32]. We briefly summarize the features below, and describe in detail below how they are collected.

1. *Key-phrase occurrence*: Using statistical relationships, and linguistic structure patterns, as described below, we identify categories of key-phrases (strongly related to the knowledge classes) that occur in the document fragment. Features are identified by grouping words together, based on semantic relatedness (see below) and syntactical relatedness. The phrases we identify are in six categories: a) Task Phrases, b) Concept Phrases, c) Coding Elements, d) Version control statements, e) directives and f) domain specific tags. The precise method identification of each key-phase is somewhat different, and is described below. For of each of these phrase categories, we count the number of times a phrase of that category occurs within a given document fragment; thus we get a *six-dimensional key-phrase category vector* for each document fragment. This feature provides a *prima facie* indication of the possible knowledge classes associated with a knowledge fragment. The following contains the study of presence of these identified keyphrases in different taxonomy classes. To make this study, we randomly selected 20 texts from each of the knowledge classes (except, the 12[th] class, which contains only 8 entries).

**Table-2**: Taxonomy Classes/Knowledge Type and identified keyphrases

| Knowledge Type | Types of Keyphrases (average occurrence %) here, we consider the keyphrase types, whose average occurrence frequency is more than 10.00%. |
|---|---|
| Functionality and Behavior | **Task Phrases(100%)**, Concept Phrases (12%) |
| Concepts | **Concept Phrases (100%)**, Task Phrases(25%) |
| Directives | Task Phrases (12%), Concept Phrases (15%), Coding Elements (20%), **directives (100%)** |
| Purpose and Rationale | **Task Phrases (50%), Concept Phrases (50%)**, Coding Elements (12%), **domain specific tags (70%)** |
| Quality Attributes and Internal Aspects | **Task Phrases (50%)**, Concept Phrases (30%), **Coding Elements(50%)**, directives (30%), |
| Control-Flow | **Task Phrases (60%)**, Concept Phrases, Coding Elements(30%), **directives(60%)** and domain specific tags (20%) |
| Structure | **Task Phrases(50%), Concept Phrases(50%),** Coding Elements(30%), **Domain specific tags(50%)** |
| Patterns | **Task Phrases (50%), Concept Phrases(50%), Coding Elements(50%)**, directives (14%) domain specific tags (40%) |
| Code Examples | **Coding Elements (100%)**, Concept Phrases (20%), Task Phrases(15%) |
| Environment | Task Phrases (20%), Concept Phrases(40%), Coding Elements (30%), **Version control statements (95%)**, domain specific tags (30%) |
| References | **Task Phrases (60%),** Concept Phrases (30%), c) Coding Elements(20%) , **domain specific tags (60%)** |
| Non-information | NaN |

Note: The table given above, (see Table-2), shows a pattern of relation between the presence of different types of

keywords and knowledge types. We use bold font face to highlight the high presence of one or more keywords in the different knowledge types. Thus, it motivates us to use this combination of features as an important feature.

2. *Semantic Relatedness:* Within each document fragment, we determine the pairwise semantic relatedness for each pair of words. Relatedness is measured, here, using normalized pointwise mutual information (*npmi*) score [29], This measure captures how related the occurrence of a pair of words are, given a large, relevant corpus. We calculate *npmi* using an appropriate corpus (either the full text of Python manuals, or relevant Wikipedia materials, as explained below), by counting word co-occurrences. gives rise to a symmetric square matrix. This matrix plays a critical role in our approach, as it gives a strong indication of when words belong together, and is used in several different ways. We compute two features based on this matrix.

    a. *Topical Breadth & Intensity*: We cluster the words according to the *npmi* distance-score. These clusters identify semantically well-connected word groups within the document unit. Higher cluster-count indicate a topically fragmented text, with higher topical breadth; the largest cluster indicates the proportion of cohesive words in the document fragment, and thus the topical intensity. We use both these measures.

    b. *Semantic Importance:* The *npmi* matrix induces a graph, with words being the nodes, and distances scored by *npmi* values. We use this to calculate an entropy measure (described below) indicative of the importance of each word in the fragment. We take the average over all the words in the document fragment, thus giving a measure of how important the words are in the given fragment, relative to their role in the entire relevant corpus.

3. *Lexical key-phrase Context:* This feature captures, in a count-vector, the parts of speech of words that surround the key-phrase extracted from the document fragment. We use a PoS tagger to tag the parts of speech of the words in the fragment. We consider the vector of PoS-tag counts *before* a key-phrase, *after* a key-phrase, and *between* key-phrases. These are calculated *per key-phrase class.* The situation is a bit complex when many key-phrases, or ones of different types occur; the details are given below.

4. *Syntactic Key-phrase Context:* In addition to the part-of-speech tags, we also consider the syntactic structure of the document fragment. Specifically, we build a dependency tree, using the Stanford dependency parser, which resolves dependencies between language elements. For each *pair of key-phrases*, kp1, and kp2, we calculate a *dependency distance* between pairs of words in kp1, and kp2. The shortest dependency distance is chosen as the distance between kp1 and kp2. This is a measure of the conceptual dependence of the phrases in the fragment on each other.

Once we have gathered all the features, we use a standard machine learning algorithm (SVM) to train eleven separate models, one for each of the (informative) classes given in Table-1, other than the last one.

We now explain the features above in more detail.

## 3.2 Calculating Features

The first step in calculating all our features is the calculation of *nmpi* matrix, so we begin with that.

### 3.2.1 Calculating npmi

Normalized Point-wise Mutual Information [9], is measure of the strength of the connection between pairs of words that may occur in a textual fragment under consideration. It is typically calculated from a large, relevant corpus. In our case, we the LUCENE text processing system to calculate it. The equation for normalized pointwise mutual information for two terms $t_i$ and $t_j$ is given below, as described in [9]:

$$npmi(t_i,t_j) = \begin{cases} -1 & if\ p(t_i,t_j) = 0 \\ \frac{\log p(t_i) + \log p(t_j)}{\log p(t_i,t_j)} - 1 & otherwise \end{cases} \quad (1)$$

Where, $p(t_i,t_j)$ = is the joint probability of the two terms occurring in proximity. This value is estimated from the corpus \ by counting the number of observations of words $t_i$ and $t_j$ in a window of size 'K' words, (K>2). $p(t_i)$ = probability of occurrence of $t_i$ in the corpus, estimated as the relative frequency of the term in the corpus.

In general, when calculating *npmi,* it is vital to use a large, and representative corpus, so that the strength of the connection between pairs of words, as estimated from the corpus is meaningful in the context of the text fragment under consideration.

### 3.2.2 Measuring topical breadth & intensity

The pairwise *npmi* measures are calculated as above for all the pairs of words in a document unit, constituting a symmetric matrix. Using this matrix, we cluster words that are semantically related together, using an agglomerative clustering algorithm [10]. This type of clustering algorithm works bottoms up, attempting to group words that are semantically close together, as indicated by the *npmi* score. The result is a set of clusters of words, each cluster representing some sort of topically coherent set.

Consider the given input text:

*"Note When passing a ZipInfo instance as the zinfo_or_arcname parameter, the compression method used will be that specified in the compress_type member of the given ZipInfo instance. By default, the ZipInfo constructor sets this member to ZIP_STORED."*

From this text, we delete the stop words[2], obtaining the text: *{passing, ZipInfo, instance, zinfo_or_arcname, parameter compression, method, used, specified, compress,_type, member, constructor, sets member, ZIP_STORED}.*

After running the clustering procedure on this bag of words, we obtain the following clusters:

*{passing, ZipInfo, ZIP_STORED, zinfo_or_arcname, compression, compress_type}*

*{parameter, method, specified, member, constructor, sets, member}*

From the clusters above, we get a cluster-count of 2 (indicating a topic-breadth of 2). Since the larger cluster has 8 words, we get an intensity measure of 0.5: (the full unit has 16 non-stop words).

---

[2] Stop words" are those words that are considered general, and not specific any given content, like prepositions, pronouns, conjunctions, etc.

### 3.2.3 Dependency Parsing

To extract key-phrases corresponding to tasks, concepts, version control statements etc., we need to be aware of both *syntactic structure* (verb phrases, adverbial phrases etc), as well as *dependency structure.* For example, consider the sentence "The boy caught the Carp with a fishing pole", with the noun phrase, "the boy", as the subject of the verb phrase "caught the Carp". Now the adverbial phrase "with the fishing pole" is also part of the *syntax* of the sentence; the attachment of the verb phrase to the varb phrase is a dependency. So also is the dependency of the object "Carp" on the verb "caught", and the dependency of "pole" on the modifier "fishing". Syntax & dependency relationships, together with *npmi,* are the two crucial elements of keyphrase extraction.

### 3.2.4 Key-Phrase Extraction

We extract 3 types of key-phrases from a document unit: Tasks, Concepts, and Version Control Statements. All three are extracted using a combination of *npmi* values (4.2.1), and dependency structures (4.2.3). The extraction process for all 3 begin with a lemmatization[3] step. "Lemmatization use the vocabulary (dictionary) and morphological analysis of words, normally aim to remove inflectional endings only and to return the base or dictionary form of a word, which is known as the lemma. For example, see text ["before"⇔ "after"] lemmantization: [*"Returns an iterable yielding all matching elements in document order"* ⇔ *"Returns an iterable yielding all match element in document order"*]. Once this step is completed, we identify *candidate phrases;* then we prune and subset the candidate phrases using *npmi* and dependency structure to identify the 3 different types of key-phrases.

**Candidate Phrase Identification:** We identify candidate phrases by selecting the words that occur strictly between punctuations and certain stop words. However, candidate phrases *do* include certain other stop words (specifically *of, for, and, in, with, to* and *on)* because these words can be part of task and concept key-phrases. However, the candidate phrases selected in this step can include noise and semantically and syntactically unrelated content; the subsequent steps, which are specific to each type of key-phrase, are designed to eliminate words unrelated to specific type of key-phrase.

**Extracting Concept key-phrases** Concept phrases, as described in [3], as "key topical terms". To extract these, we start with candidate phrases as described above, and identify the longest meaningful sequences of nouns and adjectives. For this we need two things: part-of-speech (PoS) tags, and a way to identify the strength of connection between subsequent words. For the former, we just use the Stanford PoS tagger; for the latter, we use the *npmi* score, as calculated over a corpus of Python API documentation. The extraction process is summarized below; recall while reading that *npmi* scores range from -1 to +1, and positive scores indicate semantic relatedness in the relevant corpus.

**Algorithm:**

Step 1. Start with a candidate phrase as identified in 4.2.4

Step 2. Within the sequence of words constituting the candidate phrase, find the longest sequence of *nouns and adjectives* so that each successive pair of words has a *positive npmi score.*

Step 3. Add each such word-sequence to the list of concept phrases, having phrase size greater than or equal to two (N-gram, N>=2). Ignore the other words.

For example, the text: *"The core built-in types for manipulating binary data are bytes and byte-array.",* we extract the concept phrase*: "core built-in type"*

**Extracting Task key-phrases.** Task phrases represent actions. As for concepts, we start with a candidate phrases (extracted as described above). The next step is to identify an action or task phrase. As described by Treude *et al* [2], we do this by selecting subsequence of words in the candidate phrase that a) begins with a verb, and b) is followed by a sequence of words that are syntactically, and semantically related. The syntactically related words are identified by certain action-related syntactic dependencies, using the Stanford dependency parser and the semantically related words are identified using *npmi* scores.

The Stanford dependency parser finds dependencies such as *direct object* (a noun phrase that is the object of a verb) and *prepositional modifier* (a prepositional phrase which can modify the meaning of a verb). Treude *et al* [2], in their Tables 3 and 4, identify a set of such dependencies that connect verbs to related words in a sentence. We use these, and several others found by the Stanford parser. A full list of dependencies is omitted for brevity.

**Algorithm:**

Step 1. Starting with the candidate phrase, select longest sequences which comes under one of the task-related dependencies. If no such words are found, the candidate phrase is discarded.

Step 2. Within each such sequence, search for a verb; remove sequences that do not contain verbs.

Step 3. For each remaining sequence with at least two words, collect the subsequence (starting with a verb) where each successive pair of words has a positive *npmi* score.

Step 4. Each such subsequence is a *task key-phrase.*

For example: From the text: *"Try to find a library and return a pathname"* contains task phrases: *"find a library"* and *"return a pathname".*

**Extracting Version Control Statements.** These are sequences of 2 or 3 words that relate to version control. To extract these statements, as before, we started with "seed words" that relate to version control, and used *npmi* scores to extend to a longer coherent phrase.

To find the seed words, and also the domain-specific *npmi*s scores, we first gathered a corpus from Wikipedia relevant to version control. We started with the home Wikipedia page for version control[4] and extracted all the titles and hyper-link anchors to identify set of *root topical words* and phrases, such as *checkout, commit, branch, fork, release engineer,* etc. Next, we crawled all the pages linked from the home version control Wikipedia page, excluding pages that were not sufficiently similar to the home page[5]. This left us with a corpus of 571 pages, that we considered content related to version control. We calculate *npmi* distance measures between all pairs of words in this corpus, thus deriving very specific *npmi scores.* We use these scores to identify sequences of words relating to version control: we start with one of the root topical words or phrases, and then lengthen out from here using *npmi* to find semantically related word sequences.

---

[3] http://nlp.stanford.edu/IR-book/html/htmledition/stemming-and-lemmatization-1.html

[4] https://en.wikipedia.org/wiki/Version_control

[5] We considered linked pages whose Bag-of-Words cosine similarity to the root page was less than 15% to be too dissimilar. For this similarity measure, we just used the first two paragraphs in each page, since they typically summarized the content of the page quite well.

**Algorithm:**

Step 1. We identify topical version control words in the candidate phrase; if no such exists, we discard the candidate.

Step 2. Starting with this topical content, we consider candidate proximate words pairwise with the identified content, and expand out, as long as the *npmi* score between the content and the candidate words is positive.

Step 3. When the content words can no longer be expanded, we extract the contiguous sequence of words as a VC phrase, and remove it from the candidate.

Step 4. We process the remaining words in the candidate words, starting back at Step 1 above

Example of such phrases are: *"C Version"*, *"distributed locking version"*, etc.

The next 3 types of key-phrases are quite easy to extract, simply based on meta-data (e.g., HTML tags, and special symbols).

**Extracting Code Elements** These are fragments of code that occur within the document unit being analyzed. Fortunately, in on-line documentation, these are generally indicated clearly with HTML markups (either using a constant-width font, or directly with a `<code>` tag); they can be easily extracted.

**Extracting Directives** These refer to programming language keywords, compiler commands, Unix commands, etc. We collect all programming keywords and compiler directive statement list / compiler directive glossary list from website[6],[7].

**Extracting Domain-Specific Tags.** The occurrence of domain-specific (in this case, specific to Python APIs) is one of our features. Two distinct sources are used to extract a gazetteer of these terms. First, we gather the anchors of hyperlinks of all Wikipedia articles[8], which comes under the category *"Python"*, as domain specific tags. Our next source of domain specific tags is: Stack Overflow tags: we collect all tags from topics related to "Python" and consider them as domain specific tags.

### 3.2.5 Semantic Importance

A document unit may contain several different types of key-phrases. Once we extract different types of key-phrases from a document unit (such as concept, task, coding element, or version-control key-phrases), we would like to know *how important* a role a particular kind of key-phrase plays in that document unit. Thus, if a unit has both *concept* key-phrases, and *task* key-phrases, but the *task* key-phrase is semantically more strongly connected with the rest of the unit than the *concept* key-phrases, then the unit might more likely belong to the *Function and Behavior* knowledge class than the *Concepts* knowledge class. Thus we would like to calculate a numeric score of the semantic importance of a phrase. We calculate of a key-phrase by averaging a semantic importance score (described below) over all words in that phrase.

To calculate the importance score of a word, we borrow an information-theoretic method from social networks, where it was originally used to calculate the importance of an individual in a social network. In our case, given a document unit with *n* words, we consider an *n*-node graph $G=(N,E)$, with $|E|$ edges, each consisting of a pairs of words which have *positive npmi* scores. We ignore the pairs which have negative or zero *npmi* scores. This graph $G$ can be viewed as a sort of social network of words, where each edge *npmi* score measures the strength of connection between the words; the absence of an edge indicates the absence of a positive relationship.

We then normalize the edges, so that that the weights on all edges add up to 1 (divide each edge the sum of the edge scores). The scores on the edges can then be interpreted as a discrete probability space, where each edge is an "event", and the edge score is the probability measure; We then calculate the *entropy of G, E(G),* in the usual way, as the of the expectation of the negative log probability over the "event space" of all edges. Shetty & Adibi [11], describe how to use such a network to calculate the importance of a node: we just drop the node, and all incident edges, to yield graph *G-,* with one less node than *G,* renormalize the edge-weight "probabilities", and calculate the entropy of the remaining graph *E(G-)*. The importance score for each node is the difference of *E(G)* and *E(G-)* calculated after removing that node.

To calculate the importance score of a key-phrase, we simply calculate the sum of the importance scores for all words. Given a document unit, which contains several types of key-phrases, we sum up the importance scores *for each kind of key-phrase* that occurs in that unit; thus we obtain a score for each of *Task, Concept, Version-Control, etc.,* which measures how important each type of key-phrase is in the context of that particular document unit in which it occurs.

### 3.2.6 Key-phrase Contextual Features

In addition to extracting key-phrases, as described above in Section 4.2.4, it is also important to consider the *context* of the key-phrases. Specific ways in which the key-phrases are used within a document unit, and how the key-phrases syntactically relate to each other, can help determine the specific knowledge class to which the unit belongs. This step is influenced by the use of lexical and syntactic features by [8] (even, partially).

**Lexical Context:** We consider two classes of lexical key-phrase context features: raw word count, and PoS-tag count. The *word count context* simply counts the words before, and after the key-phrase. The *PoS tag count* context counts the number of *nouns, verbs, adverbs,* and *adjectives* before and after the key-phrase. In case there are more than one kind of key-phrase (*e.g.,* both *concept* and *task* occur in the same document unit), we calculate the two lexical context features separately for each kind of key-phrase. In case a specific key-phrase type occurs more than once (*e.g.,* two separate *task* key-phrases), we add another set of counts to the above two features, considering the word counts (or, respectively, PoS tag counts) between the *first* and *last* occurrences of the keyphrase.

**Syntactic Context:** The closeness of the syntactic relationship between key-phrases is an important feature that can help identify knowledge class. We calculate the syntactic relationship using a dependency parse tree, produced by the Stanford dependency parser. See, Figure-1.

---

[6] https://en.wikipedia.org/wiki/List_of_Unix_commands

[7] https://docs.python.org/2.5/ref/keywords.html

[8] https://en.wikipedia.org/wiki/Category:Python_(programming_language)

**Sentence:** It extracts cookies from HTTP requests

**Tagged Sentence:**
It/PRP extracts/VBZ cookies/NNS from/IN HTTP/NNP requests/NNS

**Parsed-Dependency**

```
(ROOT
  (S
    (NP (PRP It))
    (VP (VBZ extracts)
      (NP (NNS cookies))
      (PP (IN from)
        (NP (NNP HTTP) (NNS requests))))))
```

**Figure-1:** Dependency between task and concept phrases (obtained from Stanford dependency parser)

We abstract the syntactic relationship between two key-phrases as the *path length in the dependence tree*. We calculate the dependency path length, i.e., $Dep\_Path\_len(F_{k1}, F_{k2})$ (i.e., minimum count of dependency to get the relation between the words occurring in the key-phrases $F_{k1}$ and $F_{k2}$. In case there is only one key-phrase, this feature is absent; if there are more than 2 key-phrases, we consider all pairs; thus the size of his feature grows quadratically with the number of key-phrases in the document unit.

Once all the features are extracted, we use a fairly straightforward learning procedure, using Support Vector Machines (SVM). The details are presented next.

## 4. TAXONOMY CATEGORIZATION

What OntoCat faces is *a multi-class classification* problem: given a document unit, classify it into one or more of several possible class. Our approach in principle readily accommodates this; but for evaluation purposes here, since our training data classifies all document unit into one class, OntoCat will assign only one label to each document unit.

There are 12 classes. However, one of them "non-information" is essentially best recognized as not containing any interesting key-phrases or associated features. So we detect by *exclusion*. Essentially, we learn 11 different *two-class* classifiers, each of which recognizes positive or negative examples of the respective class. To train each of these, for example for the *Concepts* class, we consider all examples explicitly labeled as *Concepts* class as positive examples, and *all others* (including ones labeled as positive instances other classes) as negative examples. This "one-vs-all" approach used to train 11 classifiers, $M_1$, $M_2$, $M_3$, $M_4$, $M_5$, $M_6$, $M_7$, $M_8$, $M_9$, $M_{10}$, $M_{11}$ for each of the 11 different knowledge classes.

Given the document units, and labels, we first extract all the features described above in Section 3, for all the document units. We split the labeled data into 50% training and test data (roughly 900 samples each, details below). Each of the 11 classes have different numbers of positive examples, and an SVM for each class is trained separately, finally yielding the trained models $M_1$, thru, $M_{11}$.

Once a model $M_i$ is trained, when run over an input sentence, it assigns two probability measures: one, *positive score,* that the given input unit belongs to knowledge class *i*, and the other, *negative score,* that it does not. OntoCat then uses the following procedure to perform the classification & labeling of a given input text.

**Pseudo Code**

**Input:** (1) Text Θ to classify, (2) Classification models $M_1$, $M_2$, $M_3$, $M_4$, $M_5$, $M_6$, $M_7$, $M_8$, $M_9$, $M_{10}$, $M_{11}$.

**Output:** Knowledge taxonomy class for the given text Θ.

**Algorithm:**

Step 1. For the given text Θ

Step 2. Extract the features, as used for class "$C_1$", use extracted model "$M_1$" and run the SVM classifier.

Step 3. Repeat the same process (as given in Step 2), with $C_i$ and model $M_i$ for $2 \leq i \leq 11$.

Step 4. Now, to assign the class for Θ, select the class which shows highest probability value for positive score for membership

Step 5. If Θ is scored by all models as negative membership, then we assign it to the Non-information class, class "$C_{12}$"

## 5. EXPERIMENTS

In this section we present the experimental evaluations to evaluate efficiency the OntoCat system. We also discuss the performances of our (1) automatic task extraction system, (2) concept extraction system, and (3) version control statement extraction system.

### 5.1 Details of Dataset

We use the *pydoctypes* dataset, which is the result of a study done in Freie Universität Berlin to replicate Maalej & Robillard study [3] for Python. In contrast to the original study, however, this study marked up the documentation blocks on a stretch level rather than the block level. A stretch is usually one or more sentences, sometimes less (and supposedly never crosses a paragraph boundary). The procedure is described in German in Sven Wildermann's Bachelor thesis (see http://goo.gl/6eJBOl) In particular, the knowledge type definitions and coding instructions used are found online: (See http://goo.gl/G4vkYK). The Maalej & Robillard dataset [3], was unavailable at the time of submission, although it said to become available shortly.

Table 3**: Summary of Labeled Dataset**

| Class Name | Actual Number of Entries |
| --- | --- |
| Functionality and Behavior | 481 |
| Concepts | 220 |
| Directives | 134 |
| Purpose and Rationale | 216 |
| Quality Attribute and Internal | 143 |
| Control Flow | 28 |
| Structure (C7) | 249 |
| Patterns (C8) | 87 |
| Code Examples (C9) | 139 |
| Environment (C10) | 121 |
| References (C11) | 41 |
| Non-Info (C12) | 35 |
| TOTAL | 1864 |

### 5.2 Evaluation Strategy and Metrics

As, discussed above, the *pydoctypes* dataset contains texts related to all 12 defined taxonomies of knowledge. We divide the texts related to each class of the dataset into two halves (i.e., approx. 50% for training set and 50% for test set). The two halves chosen randomly. To mitigate the risk of bias, we repeat the process of selection of texts for training and test set 5 times. To properly evaluate the quality of the devised system, we present the average count of correct and wrong classifications for each of the classes, along with the standard deviation, and also present average recall, precision, and f-score. In addition, the prior rate varies; some knowledge classes are quite common (*Functionality & Behavior* are about 25% of the data) and some quite rare (*Non-information* occurs only a handful of times). To estimate the deviation from random guessing in all cases, we use a Hypergeometric distribution to model the probability of doing as well as we did (or better), in each case assuming that the random guesser chose the same number of candidate positive labels as our trained classifier.

### 5.3 Evaluating Quality of Extracted Task

As, task phrases are mostly related to the knowledge type: "Functionality and behavior", so we randomly select 200 sentences from dataset related to this class. We tested the performance of our devised system on these 200 extracted sentences. For this, we have collected the following evaluation data from the results:

Table-4: Evaluation of task phrase identification system

| Manually annotated task phrases (out of 200 sentences) | 135 |
|---|---|
| Task phrases correctly identified by our system | 123 |
| Total number of task phrase extracted by our devised system | 131 |

Analysis of results: Recall: 0.91. Precision: 0.94, F-score: 0.92. Significance (Hypergeometric Distribution): p << 0.001.

### 5.4 Evaluating Quality of Extracted Concepts

As, concept phrases are mostly related to the knowledge type: "Concept", so we randomly select 200 sentences from dataset related to this class. We tested the performance of our devised system on these 200 extracted sentences. For this, we have collected the following evaluation data from the results:

Table-5: Evaluation of concept phrase identification system

| Manually annotated concept phrases (out of 200 sentences) | 149 |
|---|---|
| Concept phrases correctly identified by our system | 133 |
| Number of concept phrases extracted by our devised system | 142 |

Analysis of results: Precision: 0.93, Recall: 0.89, F-score: 0.91 Significance (Hypergeometric Distribution): p << 0.001

### 5.5 Evaluating Quality of Extracted Version Control Statements

As, version control statements are mostly related to the knowledge type: "Environment", so we randomly select 200 sentences from dataset related to this class. We tested the performance of our devised system on these 200 extracted sentences. For this, we have collected the following evaluation data from the results:

Table-6: Evaluation of version control statement identification system

| Manually annotated Version Control stmts (out of 200 sentences) | 141 |
|---|---|
| Version Control Stmt correctly identified by our system | 122 |
| Number of concept phrases extracted by our devised system | 132 |

**Analysis of results**: Precision: 0.92, Recall: 0.86, F-score: 0.89 Significance (Hypergeometric Distribution): p << 0.001

### 5.6 OntoCat Performance

The summary of the labeled data is presented in Table 3. We split this data into 50% training and 50% test sets. In order to ensure that some examples of each class were available both in training and test sets, we attempted to split the items in each category roughly in half. Table 3 shows the total count of the labeled data and Table 7 shows the count of data (under each of the label) used in testing the performance of the system. We repeat the experiment 5 times. Each repetition includes the selection of fresh preparation of training and test set (by dividing data into 50% of training set and 50% test set).

The results are shown in Table-7. For all categories, as indicated by the hypergeometric p-value, we performed significantly much better than random; the largest actual p-values was 0.0002, for *non-info*, the rest were negligibly small.

The best f-score was achieved was for *Code Examples*; in our dataset, these were readily detected by HTML formatting directives, so the accuracy is not surprising. As for the rest, we see a *very* strong relationship (*Spearman's Rho = 0.8, p < 0.001*) relationship between the training set size and the f-score; the significance of this relationship is remarkable given the small sample size (just 12 classes). Thus, there is every reason to believe that as more data becomes available, we will be able to improve performance. We are told that additional data will soon become available.

### 5.7 Case Study

To gain further insight, and to examine OntoCat in a different setting, we gathered 10 sentences in a straightforwardly repeatable fashion from Python documentation as follows:

We use python document[9] and select first sentence from each of the sections. Thus we have collected total 10 sentences from starting of the document. Each of these sentences were then classified by OntoCat, and the answers were checked manually. In 7 of the 10 cases, OntoCat's labeling was correct, and they were wrong in 3 cases. We describe all the cases below. In each case, we first list the input sentence, and then discuss the labeling from Ontocat, and the correctness thereof.

1. *list.append(x): Add an item to the end of the list; equivalent to a[len(a):] = [x]*. This statement was classified as *Functionality and Be*
2. *haviour*. We rated this to be correct, since it described what a method does, as described in the coding guide.
3. *The list methods make it very easy to use a list as a stack, where the last element added is the first element retrieved ("last-in, first-out")*. This was classified as *pattern*. We rated this to be correct, since the statement describes *how to use a list as a queue*. How-to statements are to be considered patterns, as per the coding guide.
4. *To add an item to the top of the stack, use append()*. This was rated as *Functionality and Behaviour*. We considered this to be ambiguous, since it would viewed as either Functionality, or as a Pattern: it both described the function of *append()*, which could be viewed as a function; but it could also be viewed as a

---

[9] https://docs.python.org/2/tutorial/datastructures.html#dictionaries

Table 7: Performance of the Automatic Classification System

| Class Name | Correct Entry Count | Number labeled as this class (Average score) $\pm SD$ | Number of correct classifications (Average score) $\pm SD$ | Precision (average) | Recall (average) | F-Score (average) | p-value (on average score) |
|---|---|---|---|---|---|---|---|
| Functionality and Behaviour | 240 | 227.4 (6.941) | 179.0 (2.449) | 0.75 | 0.75 | 0.77 | *p<<0.001* |
| Concepts | 110 | 105.0 (2.121) | 61.0 (1.949) | 0.58 | 0.55 | 0.57 | *p<<0.001* |
| Directives | 67 | 75.8 (1.643) | 51.2 (0.836) | 0.67 | 0.76 | 0.71 | *p<<0.001* |
| Purpose and Rationale | 108 | 77.0 (0.707) | 61.6 (0.547) | **0.81** | 0.57 | 0.67 | *p<<0.001* |
| Quality Attribute and Internal | 71 | 77.4 (4.335) | 40.8 (1.303) | 0.53 | 0.58 | 0.55 | *p<<0.001* |
| Control Flow | 14 | 31.0 (3.391) | 7.0 (0.707) | 0.23 | 0.5 | 0.31 | *p<<0.001* |
| Structure | 124 | 103.6 (1.673) | 79.8 (0.836) | 0.77 | 0.65 | 0.7 | *p<<0.001* |
| Patterns | 43 | 50.4 (1.673) | 29.0 (0.707) | 0.58 | 0.67 | 0.62 | *p<<0.001* |
| Code Examples | 69 | 80.0 (1.414) | 62.2 (0.836) | 0.78 | **0.9** | **0.83** | *p<<0.001* |
| Environment | 60 | 58.2 (1.483) | 36.4 (1.410) | 0.62 | 0.6 | 0.61 | *p<<0.001* |
| References | 20 | 37.6 (2.073) | 13.0 (1.000) | 0.34 | 0.65 | 0.45 | *p<<0.001* |
| Non-Info | 2 | 4.6 (0.547) | 1.40 (0.547) | 0.2 | 0.5 | 0.29 | *p<<0.001* |
| TOTAL | 928 | 928.0 (2.334) | 625.0 (1.071) | 0.67 | 0.67 | 0.67 | *p<<0.001* |

5. *pattern* on how to add an item on a stack. Since it was ambiguous, we rated this as correct.
6. *It is also possible to use a list as a queue, where the first element added is the first element retrieved ("first-in, first-out"); however, lists are not efficient for this purpose. There are three built-in functions that are very useful when used with lists: filter(), map(), and reduce().* This was rated *Functionality & Behavior*; we considered this to be *pattern,* since it essentially discusses a *how*; a possibility of using a list data structure as a queue.
7. *There are three built-in functions that are very useful when used with lists: filter(), map(), and reduce().* This was rated *pattern,* but we rated into be incorrect, since it wasn't describing the *how* of any particular goal; we considered it *Purpose & Rationale,* since it essential describes why these methods exist: they're useful operations on lists.
8. *List comprehensions provide a concise way to create lists.* This was rated as *structure,* but we thought it was a better fit for *pattern,* since it describes how to create lists. It could also be rated *function & behavior;* since neither was selected, we consider this incorrect.
9. *The initial expression in a list comprehension can be any arbitrary expression, including another list comprehension.* This was labeled *concept;* we consider this correct, since this describes something about the concept of initial expression in a list comprehension.
10. *There is a way to remove an item from a list given its index instead of its value: the del statement, this differs from the pop() method which returns a value.* This was labeled *Structure & Relationships,* which we considered correct since it describes "how the elements" del and pop() "are related" to each other (quotes from the coding manual).
11. *Python also includes a datatype for sets.* This was rated *Concept*, which we considered correct, since the statement says something about the concept Python and it's datatypes.
12. *We saw that lists and strings have many common properties, such as indexing and slicing operations.* This was labeled *Concepts,* which we considered correct, since it describes concepts of lists & strings.

Overall the accuracy of classification was 70% (7/10). However, we note here that this is a case study rather than a rigorous statistical evaluation. Thus, while this case study provides some insights into sentences and how they are classified, we could consider the performance figures obtained in the much large samples used in the cross-validation study, as presented in Table 7, to be a much more reliable and generalizable indication of actual performance of OntoCat.

## 6. RELATED WORK

There has been significant prior interest in the study of the kinds of knowledge available within API documentation. This work can be conveniently described in three general categories.

First, *the actionability of this line of research.* Maalej & Robillard [3], assert the utility of API document knowledge classification: (1) evaluating API document content (2) development of documentation patterns to facilitate and accelerate API document creation (3) creating a vocabulary for discussions surrounding document creation & curation. There exist numerous studies on APIs, their design [12], their learnability [13], and their usability [14], but studies of API documentation are not as common. Nykasa et al. [15] performed a study to assess the documentation needs for a domain-specific API, using surveys and interviews of developers.

The Second category are *studies of classes of knowledge concerning APIs*: Hou *et al.* examined 300 questions concerning two specific Swing widgets (JButton and JTree) posted on the Swing forum [16]. They then mapped the questions to the different design features of the widgets. More recently, Ko et al. observed 17 Microsoft developers for a 90 minutes' session each, studying their information needs as they perform their software engineering tasks [17], although quite detailed, they did not specifically examine API knowledge needs. Kirk et al. investigated the knowledge problems faced when trying to develop applications which extended the JHotDraw framework [18]. After collecting 209 instances of "reuse problems" from notebooks, newsgroup postings, and student assignments, the authors identified four main categories of framework reuse problems, which Maalej & Robillard [3] argue are related their 1s categories; they make similar arguments about Sillito et al's categories; Sillito et al used grounded theory analysis of developers' "think-aloud" transcripts to develop of 44 types of questions developers ask during software evolution tasks [19]; Finally, Maalej and Happel [20] and Pagano and Maalej [22] used text mining to analyze the content of informal documentation such as commit messages, work logs, and blogs. They identified information entities and granularity levels in these types of documents. Maalej & Robillard provide a detailed analysis of all this work, and argue that the categories described align well to their categories.

The Third Category is the *automatic extraction of taxonomies*: Monperrus et al.'s describes a way to extract "programming directives" from API documentation [21]. In their work, the authors describe several linguistic patterns that manifest different types of "programming directives". We share the ultimate goal of automatically detecting knowledge types in documentation by first establishing empirical evidence on its content; our work can be viewed as extension of this work to a full set of knowledge types described by Maalej & Robillard.

The study and automatic extraction of taxonomies is also an interesting area of research in NLP community. Velardi et al. [33], proposed Ontolearn algorithm, which learns both concepts and relations from scratch via the automated extraction of terms, definitions, and hypernyms. It provides a dense, cyclic and potentially disconnected hypernym graph of terms. Wu, W. et al., [34] presents probabilistic taxonomy. De Knijff, J. et al. [35] uses domain pertinence, domain consensus, lexical cohesion, and structural relevance for automatic domain specific taxonomy creation. However, most of such efforts concentrate on term level taxonomy generation. Such efforts do not exactly classify the taxonomy of knowledge of the given document (as discussed in [3], and proved its use in software engineering documentation).

Different from prior approaches, as discussed above, we concentrate in exploring more number and categories of featured phrases (see Section 3) having different semantic influences and uses. We also utilize statistical, semantic and lexical dependencies of these features phrases with entire text, their role in entire text representation and tried to map it with different knowledge categories. After a lot of study on knowledge categorization, the current approach may be the starting approach towards the atomization of entire knowledge categorization in software engineering documents.

## 7. CONCLUSION AND FUTURE WORK

OntoCat, to our knowledge, is the first attempt to automatically classify textual content of API documentation into a rich set of knowledge classes. We make use of a pre-labled training set of document extracts, and a rich set of statistically, syntactic, and lexical features to train an ensemble of classifiers. We provide evidence that the performance of OntoCat in a cross-validation setting is both stable and good. We also show the use of OntoCat with a detailed case study.

We believe that automatic classification has a number of applications, in API document analysis & improvement; for the empirical study of crowd-sourced documents in StackOverflow; for automated fact extraction and question-answering; and for creating new organizing principles & tools for existing documentation.

## 8. REFERENCES

[1] Lin, D. (1998). Automatic retrieval and clustering of similar words (PDF). 17th International Conference on Computational linguistics (COLING). Montreal, Canada. pp. 768–774

[2] Treude, Christoph, Martin P. Robillard, and Barthélémy Dagenais. "Extracting development tasks to navigate software documentation." Software Engineering, IEEE Transactions on 41.6 (2015): 565-581.

[3] Maalej, Wiem, and Martin P. Robillard. "Patterns of knowledge in API reference documentation." Software Engineering, IEEE Transactions on 39.9 (2013): 1264-1282.

[4] D. Hou, K. Wong, and J. H. Hoover, "What can programmer questions tell us about frameworks?" in Proceedings of the 13th International Workshop on Program Comprehension, 2005, pp. 87–96.

[5] M. P. Robillard and R. DeLine, "A field study of API learning obstacles," Empirical Software Engineering, vol. 16, no. 6, pp. 703–732, 2011.

[6] D. Kramer, "API documentation from source code comments: A case study of Javadoc," in Proceedings of the Conference of the ACM Special Interest Group for Design of Communication, 1999, pp. 147–153.

[7] Surdeanu, M., McClosky, D., Tibshirani, J., Bauer, J., Chang, A. X., Spitkovsky, V. I., & Manning, C. D. (2010, November). A simple distant supervision approach for the TAC-KBP slot filling task. In Proc. TAC 2010 Workshop.

[8] Mintz, Mike, et al. "Distant supervision for relation extraction without labeled data." Proceedings of the Joint Conference of the 47th Annual Meeting of the ACL and the 4th International Joint Conference on Natural Language Processing of the AFNLP: Volume 2-Volume 2. Association for Computational Linguistics, 2009.

[9] Bouma, Gerlof. "Normalized (pointwise) mutual information in collocation extraction." Proceedings of GSCL (2009): 31-40.

[10] Karypis, George, Eui-Hong Han, and Vipin Kumar. "Chameleon: Hierarchical clustering using dynamic modeling." Computer 32.8 (1999): 68-75.

[11] Jitesh Shetty,Jafar Adibi; Discovering important nodes through graph entropy the case of Enron email database;KDD '2005 Chicago, Illinois.

[12] J. Stylos, B. Graf, D. K. Busse, C. Ziegler, and R. E. J. Karstens, "A case study of API redesign for improved usability," in Proc. Symp. on Visual Languages and Human-Centric Computing, 2008, pp. 189–192.

[13] J. Stylos and B. A. Myers, "The implications of method placement on API learnability," in Proceedings of the 16th ACM SIGSOFT International Symposium on the Foundations of Software Engineering, 2008, pp. 105–112.

[14] B. Ellis, J. Stylos, and B. Myers, "The Factory pattern in API design: A usability evaluation," in Proceedings of the 29th ACM/IEEE International Conference on Software Engineering, May 2007, pp. 302–312.

[15] J. Nykaza, R. Messinger, F. Boehme, C. L. Norman, M. Mace, and M. Gordon, "What programmers really want: Results of a needs assessment for SDK documentation," in Proceedings of the 20th Annual ACM SIGDOC International Conference on Computer Documentation, 2002, pp. 133–141.

[16] D. Hou, K. Wong, and J. H. Hoover, "What can programmer questions tell us about frameworks?" in Proceedings of the

13th International Workshop on Program Comprehension, 2005, pp. 87– 96.

[17] A. J. Ko, R. DeLine, and G. Venolia, "Information needs in collocated software development teams," in Proceedings of the 29th International Conference on Software Engineering, 2007, pp. 344–353.

[18] D. Kirk, M. Roper, and M. Wood, "Identifying and addressing problems in object-oriented framework reuse," Empirical Software Engineering, vol. 12, pp. 243–274, June 2007.

[19] J. Sillito, G. C. Murphy, and K. D. Volder, "Asking and answering questions during a programming change task," IEEE Transactions on Software Engineering, vol. 34, no. 4, pp. 434–451, July-August 2008.

[20] W. Maalej and H.-J. Happel, "Can development work describe itself?" in Proceedings of the 7th IEEE International Working Conference on Mining Software Repositories, 2010, pp. 191–200.

[21] M. Monperrus, M. Eichberg, E. Tekes, and M. Mezini, "What should developers be aware of? An empirical study on the di-rectives of API documentation," Empirical Software Engineering, vol. 17, no. 6, pp. 703–737, 2012.

[22] D. Pagano and W. Maalej, "How do open source communities blog?" Empirical Software Engineering, pp. 1–35, 2012.

[23] Content Analysis for Software Documentation, Coding Guide Section, http://apidocumentation.org

[24] Ferrucci, D., Brown, E., Chu-Carroll, J., Fan, J., Gondek, D., Kalyanpur, A. A., ... & Schlaefer, N. (2010). Building Watson: An overview of the DeepQA project. *AI magazine*, *31*(3), 59-79.

[25] Blei, D. M., Ng, A. Y., & Jordan, M. I. (2003). Latent dirichlet allocation. *the Journal of machine Learning research*, *3*, 993-1022.

[26] Mintz, M., Bills, S., Snow, R., & Jurafsky, D. (2009, August). Distant supervision for relation extraction without labeled data. In *Proceedings of the Joint Conference of the 47th Annual Meeting of the ACL*

[27] Panichella, A., Dit, B., Oliveto, R., Di Penta, M., Poshyvanyk, D., & De Lucia, A.. How to effectively use topic models for software engineering tasks? an approach based on genetic algorithms. In *ICSE 2013*

[28] Linstead, E., Rigor, P., Bajracharya, S., Lopes, C., & Baldi, P. (2007, November). Mining concepts from code with probabilistic topic models. In *Proceedings of the twenty-second IEEE/ACM international conference on Automated software engineering* (pp. 461-464). ACM.

[29] Bouma, G. (2009). Normalized (pointwise) mutual information in collocation extraction. *Proceedings of GSCL*, 31-40.

[30] Cunningham, H. (2005). Information extraction, automatic. *Encyclopedia of language and linguistics,*, 665-677.

[31] Grishman, R. (1997). Information extraction: Techniques and challenges. In*Information extraction a multidisciplinary approach to an emerging information technology* (pp. 10-27). Springer Berlin Heidelberg.

[32] Min, B., Grishman, R., Wan, L., Wang, C., & Gondek, D. (2013). Distant Supervision for Relation Extraction with an Incomplete Knowledge Base. In HLT-NAACL (pp. 777-782).

[33] Velardi, P., Faralli, S., & Navigli, R. (2013). Ontolearn reloaded: A graph-based algorithm for taxonomy induction. Computational Linguistics, 39(3), 665-707.

[34] Wu, W., Li, H., Wang, H., & Zhu, K. Q. (2012, May). Probase: A probabilistic taxonomy for text understanding. In Proceedings of the 2012 ACM SIGMOD International Conference on Management of Data (pp. 481-492). ACM.

[35] De Knijff, J., Frasincar, F., & Hogenboom, F. (2013). Domain taxonomy learning from text: The subsumption method versus hierarchical clustering. Data & Knowledge Engineering, 83, 54-69.